\documentclass[12pt]{article}
\input epsf.sty

\begin{document}

\def\CA{{\cal A}}
\def\CB{{\cal B}}
\def\CC{{\cal C}}
\def\CD{{\cal D}}
\def\CE{{\cal E}}
\def\CF{{\cal F}}
\def\CG{{\cal G}}
\def\CH{{\cal H}}
\def\CI{{\cal I}}
\def\CJ{{\cal J}}
\def\CK{{\cal K}}
\def\CL{{\cal L}}
\def\CM{{\cal M}}
\def\CN{{\cal N}}
\def\CO{{\cal O}}
\def\CP{{\cal P}}
\def\CQ{{\cal Q}}
\def\CR{{\cal R}}
\def\CS{{\cal S}}
\def\CT{{\cal T}}
\def\CU{{\cal U}}
\def\CV{{\cal V}}
\def\CW{{\cal W}}
\def\CX{{\cal X}}
\def\CY{{\cal Y}}
\def\CZ{{\cal Z}}

\newcommand{\todo}[1]{{\em \small {#1}}\marginpar{$\Longleftarrow$}}
\newcommand{\pdv}[2]{{\frac{\partial #1}{\partial #2}}}
\newcommand{\labell}[1]{\label{#1}\qquad_{#1}} 
\newcommand{\ads}[1]{{\rm AdS}_{#1}}
\newcommand{\SL}[0]{{\rm SL}(2,\IR)}
\newcommand{\cosm}[0]{R}
\newcommand{\tL}[0]{\bar{L}}
\newcommand{\hdim}[0]{\bar{h}}
\newcommand{\bw}[0]{\bar{w}}
\newcommand{\bz}[0]{\bar{z}}
\newcommand{\be}{\begin{equation}}
\newcommand{\ee}{\end{equation}}
\newcommand{\lp}{\lambda_+}
\newcommand{\bx}{ {\bf x}}
\newcommand{\bk}{{\bf k}}
\newcommand{\bb}{{\bf b}}
\newcommand{\BB}{{\bf B}}
\newcommand{\tp}{\tilde{\phi}}
\hyphenation{Min-kow-ski}

\def\ie{{\it i.e.}}
\def\eg{{\it e.g.}}
\def\cf{{\it c.f.}}
\def\etal{{\it et.al.}}
\def\etc{{\it etc.}}

\def\varep{\varepsilon}
\def\del{\nabla}
\def\grad{\nabla}
\def\tr{\hbox{tr}}
\def\perp{\bot}
\def\half{\frac{1}{2}}
\def\p{\partial}

\renewcommand{\thepage}{\arabic{page}}
\setcounter{page}{1}

\rightline{Neural Comp. MS 2132}
\vskip 1cm
\centerline{\Large \bf Metabolically Efficient Information Processing}
\vskip 1cm

\renewcommand{\thefootnote}{\fnsymbol{footnote}}
\centerline{{\bf Vijay
Balasubramanian${}^{1}$\footnote{vijayb@pauli.harvard.edu},
Don Kimber${}^{2}$\footnote{kimber@sysint.org}
and
Michael J. Berry II${}^{3}$\footnote{berry@molbio.princeton.edu},
}}
\vskip .5cm
\centerline{${}^1$\it David Rittenhouse Laboratories, University of Pennsylvania,}
\centerline{\it Philadelphia, PA 19104, USA}
\vskip .5cm
\centerline{${}^2$ \it  FX Palo Alto Laboratory}
\centerline{\it 3400 Hillview Avenue}
\centerline{Palo Alto, CA 94304, USA}
\vskip .5cm
\centerline{${}^3$ \it Department of Molecular Biology,}
\centerline{\it Princeton University,}
\centerline{\it Princeton, NJ 08544, USA}
\setcounter{footnote}{0}
\renewcommand{\thefootnote}{\arabic{footnote}}

\begin{abstract}
Energy efficient information transmission may be relevant to biological
sensory signal processing as well as to low power electronic devices.
We explore its consequences in two different regimes.  In an ``immediate''
regime, we argue that the information rate should be maximized subject
to a power constraint, while in an ``exploratory'' regime, the
transmission rate {\it per} power cost should be maximized.  In the
absence of noise, discrete inputs are optimally encoded into Boltzmann
distributed output symbols. In the exploratory regime, the partition
function of this distribution is numerically equal to $1$. The structure
of the optimal code is strongly affected by noise in the transmission
channel. The Arimoto-Blahut algorithm, generalized for cost constraints,
can be used to derive and interpret the distribution of symbols for
optimal energy efficient coding in the presence of noise.  We outline
the possibilities and problems in extending our results to information
coding and transmission in neurobiological systems.
\end{abstract}

\section{Introduction: The Utility of Information}
\label{sec:util}

There is increasing evidence that far from being noisy and unreliable,
spiking neurons can encode information about the outside world
precisely in individual spike timings~\cite{rel1}, \cite{bwm1},
\cite{rel2}.  Estimates of the information transmitted by sensory
neurons have often found them to be highly informative, sending ~2 to
5 bits per spike, and quite reliable, using roughly half of the total
entropy available in their spike trains (\cite{rel2} (monkey),
\cite{inform1} (cricket), \cite{inform3} and~\cite{inform2} (frog),
[Berry et.al., 1998] (retina), \cite{inform5} (blowfly H1), \cite{inform7}
(retina), [Reinagel et.al., 1998] (cat) -- see \cite{spikesbook} for a
discussion and review.) So it is possible that there are behavioral
regimes where information theory will be a powerful tool for
predicting the structure of neural codes, provided the costs and
constraints of biological computation are properly incorporated.
Therefore, as a step towards a biologically relevant information
theory, we examine the effect of energetic costs on the coding and
transmission of information by discrete symbols, following important
prior work by Levy~\cite{levy} and Sarpeshkar~\cite{rahul}.  We have
in mind a model of a sensory system where signals from the natural
world are detected and encoded, and pass through a noisy channel
before arriving at a decision making receiver.  Our results are
equally relevant to low power electronic devices, such as mobile
telephones, that are constrained by finite battery life.

In general terms, the role of a sensory system in the process of
information use by an organism is summarized in Fig.~1.
Information about the environment is detected by sensors and encoded for
transmission through an information channel to a control system. For
example, the retina detects patterns of light, which are encoded by
ganglion cells for transmission through the optic nerve to the brain. We
might expect evolution (or engineering) to produce systems which make an
``optimal'' choice for both the {\em amount} of information to transmit,
and given the amount, for the {\em kind} of information to transmit.
The amount of information is quantified in classical information theory
by the mutual information $I(S;Z)$, and by the rate $R = I(S;Z)/N$
during a period in which $N$ symbols are transmitted ~\cite{coverbook}.
As we will describe, information theory can be used to determine the
minimum power necessary to transmit at a given rate $R$, or the minimum
energy needed to transmit a given amount $I$ of information.  The rate
at which the organism should operate is determined by a tradeoff between
the value and cost of the transmitted information.  We outline two
different behavioural regimes in which these tradeoffs leads to
different coding strategies.

\paragraph{Immediate regime: } In some activities an organism is engaged
in a time-critical task involving rapidly changing environmental
states, and its performance depends strongly on its rate of sensory
information acquisition $R$.  For example, a cheetah's effectiveness
in catching a gazelle, and hence in procuring metabolic gains from
food, might be expected to improve with increasing $R$.  However,
acquiring sensory information also incurs a metabolic cost at some rate
$E$, and unless the resulting rate of metabolic gain for the organism
$V$ is great enough, the expenditure may not be worthwhile.  While
numerous factors affect the value of information, we will focus on how
it varies with the rate $R$ and consider a value function $V(R)$ with all
other variables held constant.

In general, we expect that the value $V(R)$ of sensory information
will increase monotonically, but not linearly, with $R$.  At a low
enough rate of acquisition, the sensory modality will be of no use to
the organism.  For instance, if the cheetah sees half as well, it will
not capture half as many gazelles - it will starve.  Conversely, at a
high enough information rate the value should saturate, as there is
only so much meat in a gazelle.  Balancing the marginal increase in
value to the organism, $\partial V(R) /\partial R$, against the
marginal increase in energy expense, $\partial E(R)/ \partial R$,
yields some optimal rate $R^*$ for such an
{\em immediate regime}.    Alternatively, there may be
some structural constraints, such as signal-to-noise ratio of the
sensory modality or processing speed of the biological circuitry, that
limit the attainable rate $R^c$.

We cannot compute $R^*$ without knowing the value function $V(R)$, and
we cannot compute $R^c$ without knowing the structural constraints.
However, the smaller of these two values will set the organism's rate of
sensory information acquisition, {\em and whatever it is, an optimal
code will minimize the energy cost for this rate.}  To study the
structure of such codes we can simply ask how to minimize the power
required to transmit at a given information rate.  As we shall see,
$E(R)$ is an increasing convex function, so this is equivalent to
determining that maximum rate $R(E)$ of information transmission given a
constraint of average energy $E$ per symbol.  (See Fig.~2.)

\paragraph{Exploratory regime: }  In many other situations, the relevant
environmental state is changing slowly and an organism is not faced with
any urgent tasks.  Here, it is free to choose the rate at
which it surveys its surroundings, as well as the time it spends before
taking a behavioral decision that changes its environment.  The quality
of exploration will depend on the {\it total} amount of sensory
information acquired.  Better exploration will allow the organism to
achieve more appropriate behavior, but continued exploration will involve a
cost in metabolic energy as well as in opportunities for other
behavior.  Therefore, there will be an optimal amount of information, $I^*$,
that the organism should acquire, where the marginal value of
exploration matches its marginal cost.

We cannot compute $I^*$ without knowing the value of exploration
achievable using an amount $I$ of information.  However, {\em whatever
the value of $I^*$, and independently of the details of the activity,
an ``optimal'' sensory system will transmit that information at the
rate which minimizes the cumulative energy cost $E_c(I^*)$.} The
convexity of $E(R)$ implies that this is achieved by a sensory system
that transmits at a fixed rate, as any variations in the rate will
result in a higher cumulative energy cost.  This optimal rate of
sensory information acquisition will minimize $E_c(I^*) = E(R) {I^*
\over R}$, or equivalently, maximize $R(E)/E$.

\paragraph{Low power devices: }  Both the immediate and the exploratory
regimes apply to low power electronic devices, such as mobile telephones
or laptop computers.  The finite battery lifetime of these devices puts
a premium on energy efficiency.  The immediate regime is equivalent to
an ``on-line'' mode, where the information rate of the device is
determined by the application but the total amount of information is
variable.  The exploratory regime is equivalent to an ``off-line'' or
``batch'' mode, where the total amount of information to transmit is set,
but the rate is variable.

\paragraph{Summary: } A system operating at any given information
rate $R$ should transmit using the minimimum energy $E(R)$ required
for that rate, all other constraints being held equal.
In immediate activities the optimal rate is determined by the
tradeoff between gain realizable at rate $R$ and the cost $E(R)$.
However, in an exploratory regime, the optimal rate maximizes $R(E)/E$,
independently of the details of the activity.
The next sections describe the general structure of energy efficient codes.

\section{Metabolically constrained capacity and coding}
\label{metcap}
In this section we consider the consequences of metabolic efficiency
in information transmission.  We will not address the problem of
determining {\em what} information to transmit, but abstract the
mapping $S \rightarrow X$ in Fig.~1 as performing this
task. From this point of view, we can treat $X$ as a sequence of
symbols to be encoded into a sequence $Y$ of channel inputs,
which get transmitted to produce an output sequence $Z$.
Denote the elements of these sequences at a specific time as $x$, $y$,
and $z$.  Channel transmission is both noisy and energetically costly.

Assume a discrete memoryless channel, modeled by cross-over
probabilities $Q_{k|j} \equiv \Pr \{z=z_k | y=y_j \}$ giving the
probability that a channel input symbol $y_j$ results in a channel
output $z_k$.  The organism as a whole incurs a variety of energetic
expenditures at all times, but we will focus on the costs of operating
the sensory system, these being relevant to the optimization
considered here.  The energetic cost of transmitting information can
be referred to either the input $Y$, the output $Z$, or may even be a
function of both $X$ and $Y$.  However, we choose to associate energy
costs $\{ E_1, \cdots, E_n \}$ with input symbols $\{y_1, \cdots, y_n
\}$.  This entails no loss of generality, since for arbitrary costs
$E_{jk}$ depending on both input $y_j$ and output $z_k$ we may simply
take $E_j$ as the expected cost $E_j \equiv \sum_k Q_{k|j} E_{jk}$ for
use of symbol $y_j$.

Our goal is to find, for any given energy $E$, the maximum achievable
mutual information $I(X;Z)$ between the signal $X$ and the channel output $Z$,
with expected energy cost $\bar{E} \leq
E$.  However, it can be shown that $I(X;Z) \leq I(Y;Z)$, with equality
when $X$ can be completely determined from $Y$~\cite{coverbook}.
Intuitively, the encoding from $X$ to $Y$ should exploit the
channel characteristics, but without loss of information about
$X$. Assuming the mapping from $X$ to $Y$ is indeed lossless,
maximizing $I(X;Z)$ reduces to maximizing $I(Y;Z)$. Correlations
within the sequence $Y$ will always decrease the total amount of
transmitted information, since this is bounded above by the entropy
of $Y$.  So to maximize $I(Y;Z)$ we can assume 
that the symbols of $Y$ are independently drawn from a distribution
$q(y)$ over the channel inputs.   But both
$I(Y;Z)$ and $\bar{E}$ depend upon $q(y)$; so, formally, the problem is to determine
the function
\begin{equation}
C(E) = (1/N) \max_{q(y)~;~\bar{E} \leq E} I(Y;Z) ~~~~~;~~~~~ \bar{E} =
\sum_j q(y_j) E_j \, ,
\label{capdef}
\end{equation}
where $C(E)$ is called the {\em channel capacity-cost
function}~\cite{blahutbook}.  It is evident from (\ref{capdef}) and the 
statistical independence of symbols in $Y$ that $C(E) = R(E)$ where $R(E)$ is  the constrained transmission rate discussed earlier.   The channel coding theorems of classical information theory assert that reliable transmission
of information is possible at any rate less than $R$, and at no
rate greater than $R$.  Our focus is not on reliable transmission
{\em per se}, but simply on the maximum per symbol rate $R(E)$ at which
mutual information $I(Y;Z)$ can be established given the
constraint $\bar{E} \leq E$.

We now address, first in the noiseless case, then for a noisy
channel, the related problems of:   (1) characterizing $C(E)$,  (2)
determining the distribution $q_E(y)$ which achieves $C(E)$, and
(3) finding the maximum of $C(E)/E$.  The first two problems
are of interest because an energy-optimal device or organism should
achieve $C(E)$ for whatever energy $E$ it is operating at, requiring
a very particular distribution over $y$.
The third problem is interesting because it allows us to determine both the
rate $C^*$ and energy $E^*$ at which an energy-optimal organism would
operate in the exploratory regime, regardless of the details of its activity.

\subsection{Efficient Noiseless Transmission}

In the absence of noise, the channel input and output are equal
($Y=Z$), and the mutual information $I(Y;Z)$ equals the channel input
entropy $H(Y)$.  So, finding the capacity at fixed energy reduces to
maximizing the entropy of $Y$ at fixed energy.  Correlations within
the sequence $Y$ will always decrease the entropy, so we can assume
that the symbols in $Y$ are drawn independently from some distribution
$q$.  The purpose of the encoding process $X \rightarrow Y$ is to
implement a deterministic map between  the signal $X$ and the channel
input $Y$, 
in such a way that the symbols of $Y$ are statistically independent
and have a distribution $q$.  We will not dicuss how this encoding is
performed in practice and will focus instead on the structure of the
optimal distribution $q$.\footnote{There are standard algorithms in
coding theory that perform such mappings between $X$ and
$Y$~\cite{coverbook}.  Most such algorithms are not biologically
plausible and it would be very interesting to determine whether
suitable encoding algorithms can be implemented by biological
hardware.}  Then the per-symbol information rate (or entropy) and
energy involved in the transmission are $H = - \sum_{j=1}^n q_j \ln
q_j$ and $\bar{E} = \sum_{j=1}^n q_j E_j$, where $q_j = q(y_j)$. In
the immediate regime we maximize $H$ at fixed $\bar{E}$, while in the
exploratory regime we maximize $H/\bar{E}$.

\paragraph{Immediate regime: }
Entropy maximization at fixed average cost is a classic problem,
solvable using the method of Lagrange multipliers by defining the
function
\begin{equation}
G = - \sum_{j=1}^n q_j \ln q_j + \beta\left(\sum_{j=1}^n q_j E_j -
E\right) + \lambda\left(\sum_{j=1}^n q_j - 1 \right)
\end{equation}
and setting its derivatives with respect to $\beta$, $\lambda$, and
all the $q_j$ equal to zero.  Setting $\pdv{G}{\lambda} = 0$
ensures that the $q$ remains a probability distribution. The
conditions $\partial G/\partial q_j = \partial G/\partial \beta =
\partial G /\partial \lambda = 0$ can be solved simultaneously to
yield
\begin{equation}
q_j = {e^{-\beta E_j} \over {\cal Z}}
~~~~~;~~~~~
{\cal Z} = \sum_{j=1}^n e^{-\beta E_j}
~~~~~;~~~~~
E
= {\sum_{j=1}^n E_j e^{-\beta E_j} \over
   \sum_{j=1}^n e^{-\beta E_j} }
= - {\partial \ln {\cal Z} \over \partial \beta} \, ,
\label{boltz}
\end{equation}
where the normalization factor ${\cal Z}$ is known as the partition
function and $\beta$ is implicitly determined by demanding that the
average energy be $E$.  We are simply recovering the commonplace fact
of statistical physics that entropy is maximized at fixed average
energy by a Boltzmann distribution with an ``inverse temperature''
$\beta$ defined by (\ref{boltz}).  Standard results about Boltzmann
distributions then tell us that the maximum information rate at fixed
energy $H(E)$ is a convex function of $E$, increasing from $0$ at
$E_{{\rm min}} = \min_j (E_j)$ to a maximum $H_{{\rm max}} = \ln n$ at
$E_{{\rm max}} = \sum_{j=1}^n E_j / n$.  (In the language of
statistical physics, the ``heat capacity'' is positive.)  Larger
energies ($E > E_{{\rm max}}$) lower the entropy. (See Fig.~2.)

\paragraph{Exploratory regime: } In the exploratory regime, we
maximize the information transmitted per energy cost.   So we should
extremize
\begin{equation}
\tilde{G} = {H \over E} + \lambda\left(\sum_{j=1}^n q_j - 1\right)
= {-\sum_{j=1}^n q_j \ln q_j \over \sum_{j=1}^n q_j E_j}
+ \lambda\left(\sum_{j=1}^n q_j - 1\right)
\label{tildeg}
\end{equation}
with respect to $\lambda$ and all the $q_j$.  If $\tilde{G}$ is
maximized by some distribution $\tilde{q}$, there is a corresponding
information rate $\tilde{H}$ and power consumed $\tilde{E}$.  We have
already shown that for fixed $\tilde{E}$ the information rate is
maximized by the Boltzmann distribution (\ref{boltz}).  So $\tilde{q}$
must be Boltzmann for some inverse temperature $\tilde{\beta}$.  This
reduces the multi-variable optimization problem of maximizing
$\tilde{G}$ to a single equation -- choose $q$ to be Boltzmann as in
(\ref{boltz}) and demand that $\partial \tilde{G} / \partial \beta =
0$.  It is easy to solve this condition in terms of the partition
function (\ref{boltz}) and $H = \beta E + \ln {\cal Z}$.  Maximizing with
respect to $\beta$ gives the condition $\ln {\cal Z} \, {\partial^2 \ln {\cal Z}
\over \partial \beta^2 } = 0$. Solutions which maximize $\tilde{G}$
satisfy
\begin{equation}
\ln {\cal Z} = 0 ~~~~~~~
\Longrightarrow ~~~~~~~
{\cal Z} =1 \, .
\end{equation}
Thus, information transmission is optimized in the exploratory regime
by a Boltzmann distribution with unit partition function.  This
selects a particular energy $E^*$ and associated entropy $H^*$.
Despite the ubiquity of the partition function in statistical
physics, this is the only instance, insofar as the authors are
aware, of a clear physical meaning assigned to a particular numerical
value of ${\cal Z}$.

\subsection{Efficient Noisy Transmission}
\label{effnoisy}
Now consider the noisy channel.  Once again, the capacity will be
maximized when the symbols of the sequence $Y$ are chosen
independently from some $q(y)$ because correlations reduce
transmitted information. With this assumption, and the channel
crossover probabilities defined in Sec.~\ref{metcap}, the channel
capacity (\ref{capdef}) at a fixed transmission energy becomes
\begin{equation}
C(E) = \max_{q(y) ~;~ \bar{E} \leq E}
 \left[ - \sum _j q_j \ln
q_j + \sum_{jk} q_j Q_{k|j} \log P_{j|k}
 \right] \, ,
\label{c1}
\end{equation}
where $P_{j|k} \equiv \Pr\{ (Y=y_j | Z=z_k \}$ is given by
\begin{equation}
P_{j|k}  =  \frac{p(y=y_j,z=z_k)}{p(z=z_k)}
        ~=~ \frac{q_j Q_{k|j}}{\sum_j q_j Q_{k|j}} .
\end{equation}
The maximization is complicated by the dependency of $P_{j|k}$ on
$q_j$. An insight due to Arimoto~\cite{arimoto} and
Blahut~\cite{blahut}, which still applies despite the energy
constraint, is that (\ref{c1}) can also be written as the double
maximization
\begin{equation}
C(E) = \max_{q(y), \hat{P} ~;~ \bar{E} \leq E}
 \left[ - \sum_j q_j \ln q_j -    \hat{H}(Y|Z)
\right] \, ,
\label{ccdblmax}
\end{equation}
where we define $\hat{H}(Y|Z) \equiv \sum_j q_j \hat{H}_j \equiv - \sum_{jk}
q_j Q_{k|j} \log \hat{P}_{j|k}$.  The advantage of this form is
that the capacity can be computed numerically by an iterative
algorithm which alternately maximizes with respect to $q_j$ and
$\hat{P}_{k|j}$ while holding the other variable fixed. Each of
these maximizations can be carried out using Lagrange multipliers,
as in the previous derivations.  The resulting algorithm can be
summarized as:
\begin{enumerate}
\item Choose arbitrary nonzero $q_j^{(0)}$
\item For $t=0,1,2,...$ repeat:
\begin{enumerate}
\item
$
  \hat{P}_{j|k}^{(t)} \leftarrow
     \frac{q_j^{(t)} Q_{k|j}}{\sum_j q_j^{(t)}Q_{k|j}}
$
\item  $
            q_j^{(t+1)} \leftarrow \frac{ e^{ -\beta E_j - \hat{H}_j^{(t)}}}{
                          \sum e^{- \beta E_j -\hat{H}_j^{(t)}}}
          $
with $\beta$ chosen so $\sum_i q_j^{(t+1)} E_j = E$
\item If $q_j^{(t+1)}$ close to $q_j^{(t)}$ stop
\end{enumerate}
\end{enumerate}
The correctness of this generalization of the classic
Arimoto-Blahut algorithm  is discussed
in~\cite{blahut}. In maximizing with respect to $q$ in step (2b),
$\hat{H}(Y|Z)$ and the energy costs play identical roles.  Indeed,
$\hat{H}(Y|Z)$ is essentially the average cost due to information loss
in noise, leading to the Boltzmann distribution in step (2b).  This
algorithm yields the capacity at fixed energy $C(E)$, and the
associated distribution $q_E(y)$.  In the exploratory regime,
numerical optimization of $C(E)/E$ gives an optimal energy $E^*$,
associated capacity $C^*$ and distribution $q_{E^*}(y)$.

\paragraph{Summary: }Given the channel noise  and the symbol energies,
the capacity function $C(E)$ can be computed. In the noiseless case,
it is achieved by a Boltzmann distribution.  For a noisy channel,
$C(E)$ is computed numerically, and in all cases the distributions
produced by the algorithm above achieve metabolically optimal
transmission.  In the exploratory regime, the rate should be chosen to
maximize $C(E)/E$ which is achieved in the noiseless case when ${\cal
Z}$ equals $1$.     We have not discussed the implementation of the
encoding from $X$ into $Y$, which may be realized by either arithmetic
or block coding methods~\cite{coverbook}.  How well this mapping can
be approximated by biological organisms is a question for
investigation.

\section{Characteristics of the efficient code}
\label{effchar}

In this section, we consider some of the
properties of energy efficient codes.  First, we show that the
optimal code is invariant under certain changes in the symbol energies.
Then we illustrate some of the effects of adding noise.

\subsection{Energy invariances}
The metabolically efficient distribution on code symbols is invariant
under some transformations of the energy model in both the immediate
and exploratory regimes.   Regardless of whether the energy costs are
assigned to the channel inputs $y_i$ or the channel outputs $z_j$,
the optimal immediate symbol distributions are independent of a
constant shift in the energies ($E_k \rightarrow E_k + \Delta$).   In
the exploratory regime, the optimal distribution is independent of
rescalings of the energies ($E_k \rightarrow \lambda E_k$).   This is
shown as follows. 

\paragraph{Immediate regime: } In the immediate regime we fix the
average transmission energy ($E$), and carry out the Arimoto-Blahut
optimization algorithm in Sec.~\ref{effnoisy}.   First suppose that
symbol energies $E_j$ have been assigned to the channel inputs. We
choose an arbitrary starting distribution $q_j^{(0)}$  for the channel
inputs and iteratively perform steps (a) and (b) of the algorithm to
find improved distributions $q_j^{(t+1)}$.    Step (a) leaves
$q_j^{(t)}$ unchanged.   Step (b), which computes $q_j^{(t+1)}$, is
manifestly invariant under a constant shift of the input energies $E_j
\rightarrow E_j + \Delta$, accompanied by a shift of the average
transmission energy $E \rightarrow E + \Delta$.     So the
energy-optimal immediate distribution is invariant under a
simultaneous constant shift of all the symbol energies and the average
energy.    Next suppose that symbol costs $U_k$ have been assigned to
the channel outputs $z_k$.    The average energy expended by a channel
input $y_j$ is $E_j = \sum_k U_k \, Q_{k|j}$.       Since this
relation is linear, a constant shift by $\Delta$ of the output
energies $U_k$ translates to a constant shift by $\Delta$ of the input
energies $E_j$, leaving the optimal immediate distribution invariant.

\paragraph{Exploratory regime: }      Suppose the channel inputs have energy $E_j$ and that $C(E)$ is the channel capacity at fixed transmission energy $E$.    We compute the exploratory regime optimum by setting
\begin{equation}
{\partial (C(E)/E) \over \partial E} = {1 \over E} {\partial C(E)
\over \partial E} - {C(E) \over E^2} = 0 \, .
\label{expcond1}
\end{equation}
It follows from the Arimoto-Blahut algorithm that the optimal input
distribution at fixed transmission energy is invariant under a combined
rescaling of both the input symbol energies and the average transmission
energy ($E_k \rightarrow \lambda \, E_k$ and $E \rightarrow \lambda \, E$).
To see this, observe that step (a) of the algorithm does not change the
distribution while the condition in step (b) is solved for the new energies
by rescaling $\beta \rightarrow \beta / \lambda$.      Since the capacity
is a function of only the distribution of code symbols and not directly
of the symbol energies, we conclude that the capacity for the system
with rescaled energies, $C_\lambda$, satisfies the relation
\begin{equation}
C_\lambda(\lambda E) = C(E)  \, .
\label{capinv}
\end{equation}
To find the optimal exploratory distribution with the rescaled
energies we must solve $\partial (C_\lambda(\tilde{E})/\tilde{E}) /
\partial \tilde{E} = 0$.    Changing variables to $E =
\tilde{E}/\lambda$ and using (\ref{capinv}) we find that 
\begin{equation}
{\partial (C_\lambda(\tilde{E}) /\tilde{E})\over \partial \tilde{E}} =  {1 \over \lambda^2}  {\partial (C_\lambda(\lambda E) /E) \over \partial E} = {1 \over \lambda^2} {\partial (C(E)/E) \over \partial E} = 0
\, .
\label{expcond2}
\end{equation}
Since this equation is proportional to  (\ref{expcond1}), the optimal
exploratory distribution is invariant under a rescaling of the input
energies.     If we assign costs $U_k$ to the output symbols,
linearity of the relation $E_j = \sum_k U_k \, Q_{k|j}$ between input
and output costs implies that rescaling the output energies rescales
the effective input energies and again leaves the exploratory optimum
invariant.

\subsection{The effects of noise}

In general, an energy-efficient code should suppress the use of
expensive symbols.  However, noise can have a dramatic effect, since
conveying information requires the use of reliable symbols.  In fact,
the noisiness of a cheaper symbol can easily lead to its suppression
relative a more expensive, but reliable, symbol.  This sort of effect
is particularly important in applications to biological systems, and
is illustrated in the toy examples below.

Consider a noisy channel in which six symbols $\{y_1,\cdots y_6\}$ are
transmitted as symbols $\{z_1,\cdots z_6\}$ with channel crossover (noise)
probabilities $Q_{k|j} = \Pr\{z=z_k | y = y_j\}$ as in Sec.~\ref{metcap}.
Furthermore, let the output symbol $z_n$ have a transmission energy
of $U_n = n$.
Then the average energy of the channel input symbol
is $E_i = \sum_{n=1}^6 U_n \, Q_{n|i}$.     In the absence of any
noise at all, $Q_{k|j} = \delta_{kj}$ and so $E_i = U_i$ and the
channel input and channel output distributions for the exploratory regime are both given by:
\begin{equation}
\Pr(y_n) = \Pr(z_n)  = {e^{-\beta n} \over {\cal Z}} ~~~~~~;~~~~~~ {\cal Z} = \sum_{n=1}^6 e^{-\beta n} = 1
\end{equation}
In other words, the channel input and output distributions are both exponential and the weight in the exponential is determined by the condition ${\cal Z} = 1$.  In this case we find $\beta = 0.685$.

Next suppose that we have ``nearest neighbour noise'':
\begin{equation}
Q = \pmatrix{
1-2p    &2p &0  &0  &0  &0  \cr
p   &1-2p   &p  &0  &0  &0      \cr
0   &p  &1-2p   &p  &0  &0  \cr
0   &0  &p  &1-2p   &p  &0  \cr
0   &0  &0  &p  &1-2p   &p  \cr
0   &0  &0  &0  &2p &1-2p
}
\label{chanmat}
\end{equation}
Here $Q_{k|j}$ is the entry in the j$^{\rm th}$ row and k$^{\rm th}$
columns of the matrix $Q$.  Fig.~3 shows the optimal exploratory
regime distribution on channel output symbols, for several values of
noise parameter $p$.  Notice the marked deviation of the optimal
output distribution from a pure exponential as the noise increases.
For $p=0.25$, the least energetic symbol $y_1$, with $E_1 = 1$, is
suppressed so strongly that it is less likely than symbol $y_2$, with
$E_2 = 2$.  Among the various intricate effects we have observed in
the optimal distribution as a function of noise is a
``phase-transition-like'' behaviour where the probability of a symbol
evolves smoothly until the noise reaches some critical value, and then
drops suddenly to essentially zero.  Fig.~4 shows such effects for the
input distribution to the channel (\ref{chanmat}). 

In statistical physics, phase transitions occur due to tradeoffs
between energy and entropy.  Physical systems at finite temperature
try to minimize their energy but maximize their entropy, leading to
sharp transitions, such as the melting of ice, at a critical
temperature.  In our case, information lost to noise decreases the
mutual information between the channel input and output, and this
reduction in mutual information competes against energy minimization in the
optimization.  The sharp transitions as a function of noise (Fig.~4)
are a result of this tradeoff.  Since biological signal processing
systems are noisy, it is important for applications of our formalism
that the noise be carefully measured and included in the model.

\section{Application to neural systems}
Our primary motivation in analyzing energy efficient information
transmission is to provide a formalism which can make quantitative
predictions about the detailed structure of neural codes.  To this
end, we must identify circumstances in which the neural code can be
thought of as a sequence of discrete symbols with distinct energies.
Given such a set of symbols as well as a characterization of their
transmission noise and energy cost, we can predict the unique symbol
distribution that maximizes information transmitted per unit metabolic
energy and compare this against the measured symbol distribution. 

The vertebrate retina provides a particularly good example.  Its input
is a visual image projected by the optics of the eye; its output
consists of easily-measured action potentials.  The optic nerve, which
connects the eye with the brain, represents the visual world with many
fewer neurons than at any other point in the visual pathway,
suggesting that principles of efficient coding may be relevant.  In
addition, patterns of light with particular behavioral importance, for
instance the image of a tiger, are distributed over many photoreceptor
cells, the primary light sensors of the retina.  This makes it
difficult for any single retinal neuron to evaluate the behavioral
significance of an overall image.  Therefore, we expect that the value
of the signal transmitted by a given optic nerve fibre is closely
related to its information content in bits.

Previous studies~\cite{bwm1,bwm2} have shown that ganglion cells, the
output neurons of the retina whose axons form the optic nerve, often
transmit visual information to the brain using a discrete set of
coding symbols.  In these experiments, the retina was stimulated with
a wide variety of temporal and spatial patterns of light drawn from a
white noise ensemble~\cite{bwm1}.  Under these stimulus conditions,
ganglion cells responded with discrete bursts of several spikes
separated by long intervals of silence.  The reproducibility of these
firing events was very high: the timing of the first spike jittered by
$\sim 3$ ms from one stimulus trial to the next and the total number
of spikes varied by $\sim 0.5$ spike.  This precision implies that
each event is highly informative and that events with different
numbers of spikes can reliably represent different stimulus patterns.
In addition, correlations between successive firing events were very
weak, implying that each firing event is an independent coding symbol
that carries a discrete visual message.

This suggests that the size of each firing event (i.e., the number of
spikes it contains) may be treated as a discrete symbol $N$ in the
retinal code.  A short duration of silence may likewise be discretized
to a symbol $0$.  The experimentally measured sequence of retinal
ganglion cell events, discretized in this manner, is represented in
our model as the output sequence $Z$.  In addition, $S$ is the visual
stimulus to the retina, $X$ is output of the photoreceptors, and $Y$
is an internal retinal variable representing the ideal retinal output
prior to the addition of noise.  Repeated presentations of the same
stimulus produces a distribution of ganglion cell events with a sharp
peak at a certain symbol, and a width that we attribute to noise.
Interpreting the peak of the distribution as the intended noiseless
output $Y$, the distribution of actual ganglion cell outputs yields
the channel noise matrix required by our model.  Given a measurement
or an estimate for the energy consumption by events of different sizes
(see below), our framework then predicts a specific optimal
distribution of event sizes.  Comparison of this distribution against
the experimentally measured event distribution is a quantitative check
of the relevance of metabolically efficient coding to the retina.

More generally, our methods may be applied in any system where a
suitable discretization of the neural code is available, along with a
description of noise and costs.  The all-or-nothing character of
action potentials makes such discretization possible: by choosing an
appropriate time bin, a spiking neuron's activity becomes a sequence
of integer spike counts.  The choice of time bin and
independent ``codewords'' will depend on the neuron being studied.
The noise can be measured experimentally by repetition of an identical
stimulus and observation of the resulting distribution of output
symbols.

The symbol energy is more difficult to access experimentally.
However, Siesjo [Siesjo, 1978] and Laughlin et.al.~\cite{laughlin1}
have argued that the dominant energy cost for a neuron arises in the
pumps that actively transport ions across the cell membrane.  If this
is true, then the symbol energy can be found by simulating the known
ionic currents in a neuron to find the total charge transported during
different time periods, as this charge flow must be reversed by active 
transport in order to maintain equilibrium.  Because ionic currents are large during an
action potential, the symbol energy is likely to be given by a
baseline metabolic cost plus an additional increment per spike, $E_N =
1 + b \, N$, where $b$ is the ratio of spiking cost to baseline cost
during the time bin.  The baseline cost has components due to leak
currents, synaptic currents and other cellular metabolism.  Estimates
of $b$ vary, and depend on the neuron in question. 
While a variety of measurements indicate that electrical activity accounts for roughly
half of the brain's total metabolism~\cite{siesjo}, 
the parameter $b$ may still be small.  In any case, since cellular metabolism is 
difficult to estimate, and because it is unclear in the present context whether 
pre-synaptic and post-synaptic costs should be bundled into the expense of 
producing a spike, $b$ can be treated as a free parameter for each 
neuron, and varied to find the energy-efficient code that best agrees 
with the neuron's distribution of coding symbols.

Direct determination of metabolic activity is possible for an entire
tissue by measurements of oxygen consumption or heat production.
Furthermore, the metabolic activity of a single neuron could be obtained
by measuring the uptake of a radioactively-labeled metabolic precursor,
such as glucose, during stimulation of the neuron at different firing
rates.  Such measurements could fix or place bounds on the possible
values of $b$.    

\paragraph{Summary:} We have outlined how the formalism developed in this paper
can be applied to real neurons, with particular emphasis on retinal
ganglion cells.  Discrete output symbols may be defined by counting
the number of spikes produced within a fixed time window.  The noise
in each symbol can be experimentally measured, and the energy cost can
be estimated.  Finally, the optimal distribution of spike counts in a
symbol can be computed using our methods and compared to the actual
distribution used by the neuron.  Such a test would determine whether
the metabolic cost of information transmission is an important
constraint in the structure of a neural code.

\section{Discussion}

We have described energy efficient codes in two different regimes: an
immediate regime, where a system's {\it rate} of information transmission is
set by external constraints, and an exploratory regime, where the
total {\it amount} of information transmission is set by external
constraints.  The optimal codes in these cases are closely related,
both following a Boltzmann distribution in the symbol energies, $p_j \sim 
e^{-\beta E_j}$, when there is no noise.  In the immediate regime,
the inverse temperature, $\beta$, is set to yield the imposed
information rate, while in the exploratory regime, $\beta$ is set to make
the partition function, ${\cal Z}$, equal to one.  With the addition of
noise, the optimal code must be obtained numerically, but can always
be found using a straight-forward iterative scheme.

In delineating the immediate and exploratory regimes, we do not expect
that all of an organism's behavior can be neatly assigned to one or
the other category.  Instead, we propose here that they apply to {\em
some} behaviors.  We have argued for an immediate regime in which the
transmission rate is set by the need to respond rapidly to
environmental pressures.  However, there will certainly also be
situations where the rate is determined instead by complex
interactions involving the internal needs and constraints of the
organism.

There are also subtleties in identifying regimes of behaviour that are
``exploratory''.  We have described an idealized situtation where an
organism acquires a certain amount of sensory information before
executing a single behavior.  More realistically, the organism
simultaneously acquires sensory information relevant to many possible
behaviors, and the interplay between sensation and behavior is
ongoing.  This can be analyzed within our framework by determining the
different amounts of optimal information $I^*$ associated with each
behaviour, and then requiring that the total amount of data be gathered
simultaneously.  The exploratory regime optimization continues to
determine the total rate at which the information should be gathered.  The
essential point is that in this regime the organism's behavior is
open-ended: it has sufficient time to choose a rate of sensory
information acquisition that achieves energy efficiency, while still
being able to acquire enough information to make a ``good'' behavioral
decision among the available choices. 

We have described how our formalism can be applied to a biological
system, like the retina.  Our methods should also be useful in the
analysis of low power engineered systems, such as mobile telephones or
laptop computers which use discrete, independent coding symbols.
In this case, the engineer controls the particular choice of coding
symbols, as well as the design of the encoding algorithm and the
transmission channel.  The energy and noise characteristics of the
channel can therefore be precisely determined as inputs to our
theoretical analysis.  Perhaps such an exercise will help in designing
low power devices that can perform for longer times before running
down their batteries.

\paragraph{Acknowledgements: } M.B. was supported by  a National Research
Service Award from the National Eye Institute.   V.B. was initially supported by the
Harvard Society of Fellows and the Milton Fund of Harvard University. V.B.
and D.K. are grateful to the Xerox Palo Alto Research Center, the Institute
for Theoretical Physics at Santa Barbara and the Aspen Center for Physics
for hospitality at various stages of this work.

After this work was completed, we became aware that some of the results 
presented here have been obtained independently by Gonzalo Garcia de
Polavieja.


\newpage

\begin{center}
{\bf Captions}
\end{center}

\paragraph{Fig. 1: } Schematic view of an information system.

\paragraph{Fig. 2: } Schematic of energy optimization.  The information rate
(thick line) is a convex function of the energy rate until $E_{\rm max}$.
The exploratory regime optimum ($R^*$, $E^*$) is given by the intersection of
the tangent from the origin (thin line) with $R(E)$.    

\paragraph{Fig. 3: } The effects of noise.  Probability distribution
of channel output symbols as a function of increasing nearest
neighbour noise.  The values of $p$ and the associated optimal $\beta$
displayed above are $\{p=0,\beta=0.685\}$, $\{p=0.1,\beta =0.420\}$,
$\{p=0.2, \beta = 0.340\}$, and $\{p=0.25, \beta = 0.317 \}$.

\paragraph{Fig. 4: } Sharp transitions in symbol probabilities due to noise.
Shown here is the probability of channel input symbols as a function
of noise.  Top row, left to right: $y_1$, $y_2$, $y_3$; bottom row,
left to right: $y_4$, $y_5$, $y_6$.  Notice the different vertical
scales in each panel.


\begin{thebibliography}{de Ruyter et.al., 1997}

\bibitem[Arimoto, 1972]{arimoto}
S.~Arimoto.
\newblock An algorithm for computing the capacity of an arbitrary discrete  memoryless channel.
\newblock {\em IEEE Trans. on Info. Theory}, IT-18:14--20, 1972.


\bibitem[Berry et.al., 1997]{bwm1}
M.J.~Berry II, D.W.~Warland, and M.~Meister.
\newblock The structure and precision of retinal spike trains.
\newblock {\em Proc. Natl. Acad. Sci. USA}, 94:5411--5416, 1997.


\bibitem[Berry et.al., 1998]{inform4}
M.J.~Berry II and M.~Meister.
\newblock Refractoriness and neural precision.
\newblock Journal of Neuroscience, 18:2200-2211, 1998

\bibitem[Berry et.al., 2000]{bwm2}
M. J. Berry II, D. W. Warland, and M. Meister. 
\newblock Firing events: fundamental symbols in the retinal code. 
\newblock In preparation.


\bibitem[Blahut, 1972]{blahut}
R.E.~Blahut.
\newblock Computation of channel capacity and rate distortion functions.
\newblock {\em IEEE Trans. on Info. Theory}, IT-18:460--473, 1972.

\bibitem[Blahut, 1987]{blahutbook}
R.E.~Blahut.
\newblock {\em Principles and {P}ractice of {I}nformation {T}heory}.
\newblock Addison-Wesley, Massachusetts, 1987.

\bibitem[Buracas et.al., 1998]{rel2}
G.T.~Buracas, A.M.~Zador, M.R.~deWeese, and T.D.~Albright.
\newblock Efficient discrimination of temporal patterns by motion-sensitive neurons in the primate visual cortex.
\newblock {\em Neuron}, 20(5):959-969, 1998.



\bibitem[Cover et.al., 1991]{coverbook}
T.M.~Cover and J.A.~Thomas.
\newblock {\em Elements of {I}nformation {T}heory}.
\newblock Wiley, New York, 1991.

\bibitem[de Ruyter et.al.,1997]{rel1}
R.R. de~Ruyter~van Steveninck, G.D.~Lewen, S.P.~Strong, R.~Koberle, and  W.~Bialek. \newblock  Reproducibility and variability in neural spike trains.
\newblock {\em Science}, 275:1805--1808, 1997



\bibitem[Laughlin et.al., 1998]{laughlin1}
S.B.~Laughlin, R.~de~Ruyter~van Steveninck, and J.C.~Anderson.
\newblock The metabolic cost of neural information.
\newblock {\em Nature Neuroscience}, 1(1):36--41, 1998.


\bibitem[Levy et.al., 1996]{levy}
W.B.~Levy and R.A.~Baxter.
\newblock Energy-efficient neural codes.
\newblock {\em Neural Computation}, 8:531--543, 1996.


\bibitem[Reinagel et.al., 1998]{inform8}
P.~Reinagel and R.C.~Reid.
\newblock Visual stimulus statistics and the reliability of spike
timing in the LGN.
\newblock Soc. Neurosci. Abstr. 24:139, 1998.

\bibitem[Rieke et.al., 1993]{inform3}
F.~Rieke, D.~Warland and W.~Bialek.
\newblock Coding efficiency and information rates in sensory neurons.
\newblock Europhys. Lett. 22:151-156, 1993.


\bibitem[Rieke et.al., 1995]{inform2}
F.~Rieke,  D.~Bodnar and W.~Bialek.
\newblock Naturalistic stimuli increase the rate and efficiency of
information transmission by primary auditory neurons.
\newblock Proc. Royal Society of London Series B, 262:259-265, 1995.


\bibitem[Rieke et.al., 1997]{spikesbook}
F.~Rieke, D.~Warland, R.R.~de Ruyter van Steveninck.
\newblock Spikes: Exploring the neural code.
\newblock MIT Press, Cambridge, Mass., U.S.A., 1997.


\bibitem[Sarpeshkar, 1998]{rahul}
R.~Sarpeshkar.
\newblock Analog versus digital: extrapolating from electronics to
neurobiology.  
\newblock {\em Neural Computation}, 10:1601--1638, 1998.


\bibitem[Siesjo, 1978]{siesjo}
B.K.~Siesjo.
\newblock Brain energy metabolism.
\newblock John Wiley and Sons, New York, U.S.A., 1978.

\bibitem[Strong et.al., 1997]{inform5}
S.P.~Strong, R.~Koberle, R.R.~de Ruyter can Steveninck, W.~Bialek.
\newblock Entropy and information in neural spike trains.
\newblock Phys. Rev. Lett. 80:197-200, (1997).

\bibitem[Warland, 1991]{inform1}
D.~Warland.
\newblock Reading between the spikes: real-time processing in neural systems.
\newblock Dissertation, University of California at Berkeley, 1991.


\bibitem[Warland et.al., 1997]{inform7}
D.K.~Warland, P.~Reinagel and M.~Meister. 
\newblock Decoding visual information from a population of retinal
ganglion cells.
\newblock J. Neurophysiol. 78(5):2336-2350, 1997.


\end{thebibliography}
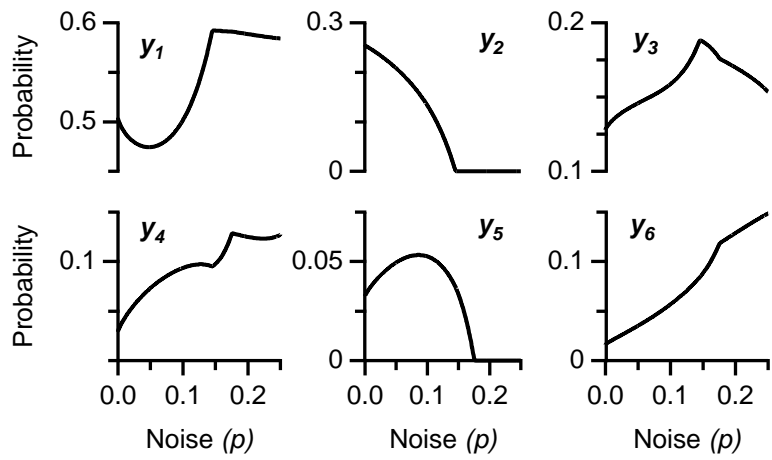
\end{document}